\documentclass{ws-p9-75x6-50}
\def \Bar#1 { \overline{#1} }

\def \eps {\varepsilon}
\def \BE {\begin{equation}}
\def \EE {\end{equation}}
\def \BEAH {\begin{eqnarray*}}
\def \EEAH {\end{eqnarray*}}
\def \BEA {\begin{eqnarray}}
\def \EEA {\end{eqnarray}}
\def \BDM {\begin{displaymath}}
\def \EDM {\end{displaymath}}
\def \na {\alpha}

\def \nb {\beta}

\def \ng {\gamma}

\def \ne {\varepsilon}

\def \nl {\lambda}

\def \nm {\mu}

\def \nn {\nu}

\def \nr {\rho}
\def \ncr {\bar \rho}
\def \ns {\sigma}

\def \nt {\tau}

\def \d {\delta}

\def \dsu#1#2 {\nabla^{ {\mbox{ {\tiny $\!\!\!\! #1 \!\dot #2 $}\rm}}} }
\def \dsd#1#2 {\nabla_{\mbox{{\tiny $\!\!\! #1 \!\dot#2 $}\rm}}}
\def \sou#1 {o^{{\mbox{{\tiny $ #1 $}\rm}}}}
\def \sod#1 {o_{{\mbox{{\tiny $ #1 $}\rm}}}}
\def \csou#1 {{\bar o}^{{\mbox{{\tiny $ \dot #1 $}\rm}}}}
\def \csod#1 {{\bar o}_{{\mbox{{\tiny $ \dot #1 $}\rm}}}}
\def \siu#1 {\iota^{{\mbox{{\tiny $ #1 $}\rm}}}}
\def \sid#1 {\iota_{{\mbox{{\tiny $ #1 $}\rm}}}}
\def \csiu#1 {{\bar \iota}^{{\mbox{{\tiny $ \dot #1 $}\rm}}}}
\def \csid#1 {{\bar \iota}_{{\mbox{{\tiny $ \dot #1 $}\rm}}}}
\def \om#1  { {\bf \omega^{ \!\!\! {\hat {\mbox{ {\tiny #1 } } } }}} \rm}
\def \com#1  { {\bf {\bar \omega}^{ \!\!\! {\hat {\mbox{ {\tiny #1 } } } }}} \rm}

\def \ospu#1#2 {#1^{ {\mbox{ {\tiny $\!\!\! #2 $} \rm}}} }
\def \ospd#1#2 {#1_{ {\mbox{ {\tiny $\!\!\! #2 $} \rm}}} }
\def \ospud#1#2#3 {#1^{ {\mbox{ {\tiny $\!\!\! #2 $} \rm}}}
                     _{ {\mbox{ {\tiny $\!\!\! \ \ #3 $} \rm}}}}
\def \ospdu#1#2#3 {#1_{ {\mbox{ {\tiny $\!\!\!  #2 $} \rm}}}
                     ^{ {\mbox{ {\tiny $\!\!\! \ \ #3 $} \rm}}}}
\def \SSu#1#2  {\mathop{S^{\scriptscriptstyle #2}}
\limits_{\scriptscriptstyle [#1] \hfill}}
\def \SSd#1#2  {\mathop{S_{\scriptscriptstyle #2}}
\limits_{\scriptscriptstyle [#1] \hfill}}
\def \cSSu#1#2  {\mathop{{\bar S}^{\scriptscriptstyle #2}}
\limits_{\scriptscriptstyle [#1] \hfill}}
\def \cSSd#1#2  {\mathop{{\bar S}_{\scriptscriptstyle #2}}
\limits_{\scriptscriptstyle [#1] \hfill}}

\def \Su#1#2 {S^{ {\mbox{ {\tiny $\!\!\! #1 $} \rm}}}_{#2}}
\def \Sd#1#2 {S_{ {\mbox{ {\tiny $\!\!\! #1 $} \rm}}}^{#2}}

\def \P#1 {\Psi_{#1}}
\def \epsu#1 {\eps^{ {\mbox{ {\tiny $\!\!\! #1 $}\rm}}} }
\def \epsd#1 {\eps_{ {\mbox{ {\tiny $\!\!\! #1 $}\rm}}} }
\def \t#1 { \dot #1 }
\def \skd#1#2  {\delta^{ {\mbox{ {\tiny $\!\!\! #1 $} \rm}}}
                      _{ {\mbox{ {\tiny $\!\!\! #2 $} \rm}}} }
\def \kd#1#2 {\delta^{#1}_{#2}}
\def \sgu#1#2  {\sigma^{ {\mbox{ {\tiny $\!\!\! #1 $} \rm}}}
                    _{ #2 }}
\def \csgu#1#2  {{\bar \sigma}^{ {\mbox{ {\tiny $\!\!\! #1 $} \rm}}}
                    _{ #2 }}

\def \sgd#1#2  {\sigma_{ {\mbox{ {\tiny $\!\!\! #1 $} \rm}}}
                    ^{  #2 }}
\def \otud#1#2#3 { #1 ^{#2}_{\ #3}}
\def \otdu#1#2#3 { #1 _{#2}^{\ #3}}
\def \ospsmuu#1#2#3 {#1^{{\cal #2 }
                     { {\mbox{ {\tiny $\!\!\! #3 $} \rm}}}}}
\def \ospsmdd#1#2#3 {#1_{{\cal #2 }
                     { {\mbox{ {\tiny $\!\!\! #3 $} \rm}}}}}
\def \ospsmud#1#2#3 {#1^{\cal #2 }
                       _{\mbox{ {\tiny $ \ #3 $} \rm}}}
\def \ospsmdu#1#2#3 {#1_{\cal #2 }
                       ^{\mbox{ {\tiny $ \ #3 $} \rm}}}
\def \ospsmddu#1#2#3#4 {#1_{{\cal #2 }{\mbox{ {\tiny $ \!\! #3 $} \rm}}}
                       ^{\mbox{ {\tiny $ \ \ \ \ #4 $} \rm}}}

\begin{document}

\title{Curvature invariants in algebraically special spacetimes }

\author{V. Pravda}

\address{Mathematical Institute, Academy of Sciences, \v Zitn\' a 25,  115 67 Prague 1, Czech Republic \\E-mail: pravda@math.cas.cz}
\author{J. Bi\v c\' ak}
\address{Institute of Theoretical Physics, Faculty of Mathematics and Physics, Charles~University, V Hole\v sovi\v ck\' ach 2, 180 00, Prague 8, Czech Republic
 \\ E-mail: bicak@mbox.troja.mff.cuni.cz}  

\maketitle

Let us define a curvature invariant of the order $k$ as a scalar polynomial constructed from $g_{\alpha \beta}$, the  Riemann tensor $R_{\alpha \beta \gamma \delta}$, and covariant
derivatives of the Riemann tensor up to the order $k$. According to this definition, the Ricci curvature scalar $R$ or the Kretschmann curvature scalar $R_{\alpha \beta \gamma \delta}R^{\alpha \beta \gamma \delta}$ are 
 curvature invariants of the order zero and $R_{\alpha \beta \gamma \delta ; \varepsilon}R^{\alpha \beta \gamma \delta  ; \varepsilon}$ is a curvature invariant of the order 1. 

We consider only vacuum spacetimes  so that the Riemann tensor is equal to the  Weyl tensor. An arbitrary curvature invariant can thus be expressed in terms of the Weyl spinor $\ospd{\Psi}{ABCD} $ and its covariant derivatives. 

We can use a standard basis
 {$\sod{A} $, $\sid{A} $}, which satisfies
\BE
\sod{A} \siu{A} =1,\quad \sod{A} \sou{A} =0, \quad \sid{A} \siu{A} =0, \label{basis}
\EE
 to decompose the Weyl spinor in the form
\BEA
\ospd{\Psi}{ABCD} =&& \Psi_0 \sid{A} \sid{B} \sid{C} \sid{D} 
                                     - 4 \Psi_1 \sod{(A} \sid{B} \sid{C} \sid{D)}
                                     + 6 \Psi_2 \sod{(A} \sod{B} \sid{C} \sid{D)} \nonumber \\
                                    &&-  4 \Psi_3 \sod{(A} \sod{B} \sod{C} \sid{D)}
                                     + \Psi_4  \sod{A} \sod{B} \sod{C} \sod{D}  \label{gendec}   \ .
\EEA
We concentrate on spacetimes of the Petrov type-{\it{III}} and {\it{N}}.
Since three principal spinors of $\ospd{\Psi}{ABCD} $  coincide  in type-{\it{III}} spacetimes,  it is convenient to choose this repeated principal spinor as the basis spinor $\sod{A} $. Then
\BE
\ospd{\Psi}{ABCD} = \sod{(A} \sod{B} \sod{C}  \ospd{\delta}{D)} \ ,
\EE 
and
\BE
\Psi_0 = \Psi_1 =\Psi_2 = 0 \ . \label{NPV0}
\EE
The Weyl spinor has the form
\BE
\ospd{\Psi}{ABCD} =
                                    -  4 \Psi_3 \sod{(A} \sod{B} \sod{C} \sid{D)}
                                     + \Psi_4  \sod{A} \sod{B} \sod{C} \sod{D}  \ .   \label{rozkweyl3}
\EE
In type-{\it N} spacetimes, all four principal spinors of $\ospd{\Psi}{ABCD} $  coincide and thus 
also $\Psi_3$ vanish.

Now it is obvious that all curvature invariants of the order zero vanish for type-{\it{III}} and type-{\it{N}} vacuum
spacetimes as a consequence of the relation $\sod{A} \sou{A} =0$. The question  arises, whether there exist
some non-vanishing curvature  invariants of higher orders.  Let us  summarize main results -  corresponding proofs and
calculations can be found in the papers\cite{BiPraN,Pra} for type-{\it{N}} 
and type-{\it{III}} spacetimes, respectively.

It turns out that the results depend  on whether the Newman-Penrose coefficient $\rho=\theta + i \omega$ vanishes or not (i.e., whether the corresponding spacetime has a non-vanishing expansion $\theta$ or twist $\omega$). 

For  $\rho=0$ the following can be shown:\\[1mm]
{\bf Proposition}\\
{\it In type-{\it N} or type-{\it III} vacuum spacetimes, admitting a
non-expanding and non-twisting null geodesic congruence,
all curvature invariants constructed  from the Riemann tensor and its
covariant derivatives of arbitrary order vanish.} 

In the case with $\rho \not=0$, all curvature invariants of the first order in type-{\it N} vacuum spacetimes vanish but  there exists a non-vanishing curvature invariant
of the first order in type-{\it III} vacuum spacetimes, which, in appropriately chosen spinor basis,  can be expressed as follows:
\BE
\ R^{\na \nb \ng \d ; \ne } R_{\na \nm \ng \nn ; \ne } R^{\nl \nm \nr \nn ; \ns }
R_{\nl\nb\nr\d;\ns } = (48 \rho \bar \rho \Psi_3 \bar  \Psi_3)^2 \ .
\EE
In the type-{\it N} vacuum spacetimes with expansion or twist (i.e., $\rho \not=0$), the first
non-vanishing curvature invariant is of the second order;
\BE
 R^{\na \nb \ng \d ; \ne \phi} R_{\na \nm \ng \nn ; \ne \phi} R^{\nl \nm \nr \nn ; \ns \nt}
R_{\nl\nb\nr\d;\ns\nt} =( 48 \nr^2 \ncr^2 {\Psi_4} {\bar \Psi_4} )^2 . \label{InvN}
\EE
Properties of curvature invariants for different Petrov types are summarized in Table 1.

The non-vanishing invariant (\ref{InvN}) can be used to show that twisting and diverging
type-{\it N} solutions analyzed recently\cite{Finley} cannot represent smooth radiation fields outside
bounded sources. 
Solutions with invariants of all orders vanishing may be of importance from quantum perspective\cite{Gibbons}.

\begin{table}{{\bf Table 1}: Curvature invariants in  vacuum spacetimes (0 - vanish, 1 - do not vanish)  }\\[1ex]
  \begin{tabular}{|c|c|c|c|c|c|}\hline
     Petrov type                                        &  I, II, D    & III     & III     &  N    & N     \\ \hline
     expansion and twist                          &     & $\nr \not= 0$     & $\nr =0$     & $\nr \not =0$     & $\nr =0 $     \\ \hline
     curvature invariants of order 0       &    1  & 0     & 0     &  0    & 0     \\ \hline
     curvature invariants of order 1       &     1  &  1   & 0     &  0    & 0     \\ \hline
     curvature invariants of order 2       &     1  &  1    & 0     &  1    & 0     \\ \hline
     curvature invariants of order $>2$ &  1    &   1    &  0    &   1   &  0    \\ \hline
  \end{tabular} \\[0.5ex]
{}
\end{table}
\section*{Acknowledgment}
V. P. acknowledges support from  grants \mbox{{GA\v CR} 201/98/1452},
\mbox{{GA\v CR} 202/00/P030.}


\begin{thebibliography}{99}

\bibitem{BiPraN} J. Bi\v c\' ak, V. Pravda, \Journal{\em Class.  Quantum  Grav.}{15}{1539}{1998}.

\bibitem{Pra}V. Pravda, \Journal{\em Class.  Quantum  Grav.}{16}{3321}{1999}.

\bibitem{Finley} J. D. Finley, J. F. Pleba\' nski, M. Przanowski, \Journal{\em Class.  Quantum  Grav.}{14}{487}{1997}.
\bibitem{Gibbons} G. W. Gibbons, \Journal{\em Class.  Quantum  Grav.}{16}{L71}{1999}.
\end{thebibliography}
\end{document}